\documentstyle[11pt,aaspp4,tighten]{article}
\def\etal{{\it et~al.\ }}
\def\eg{{\it e.g.\ }}
\def\ie{{\it i.e.,\ }}

\begin{document}
  
\title{The Parker Instability in a Thick Galactic Gaseous Disk:
I. Linear Stability Analysis and Nonlinear Final Equilibria}
\author{Jongsoo Kim\altaffilmark{1}, 
        Jos\'e Franco\altaffilmark{2},
        S. S. Hong\altaffilmark{3},
        Alfredo Santill\'an\altaffilmark{2,4}
        and Marco A. Martos\altaffilmark{2}}

\altaffiltext{1}
{Korea Astronomy Observatory, 61-1, Hwaam-Dong, Yusong-Ku, Taejon 305-348, 
Korea; \\ jskim@hanul.issa.re.kr}
\altaffiltext{2}
{Instituto de Astronom\'\i a, Universidad Nacional Aut\'onoma 
de M\'exico, A. P. 70-264, 04510 M\'exico D.F., M\'exico;
pepe@astroscu.unam.mx, marco@astroscu.unam.mx}
\altaffiltext{3}
{Department of Astronomy, Seoul National University, Seoul 151-742, 
Korea; sshong@astroism.snu.ac.kr}
\altaffiltext{4}
{C\'omputo Aplicado--DGSCA, Universidad Nacional Aut\'onoma de
M\'exico, A. P. 70-264, 04510 M\'exico D.F., M\'exico;
alfredo@astroscu.unam.mx}

\begin{abstract}
A linear stability analysis of a multi-component and magnetized Galactic disk 
model is presented. The disk model uses the observed stratifications for the 
gas density and gravitational acceleration at the solar neighborhood and, in 
this sense, it can be called a realistic model. The distribution of the total 
gas pressure is defined by these observed stratifications, and the gaseous disk 
is assumed isothermal. The initial magnetic field is taken parallel to the 
disk, with a midplane value of 5 $\mu$G, and its stratification along the 
$z$-axis is derived from the condition of magnetohydrostatic equilibrium in an 
isothermal atmosphere. The resulting isothermal sound speed is $\sim 8.4$ km 
s$^{-1}$, similar to the velocity dispersion of the main gas components within 
1.5 kpc from midplane. The thermal-to-magnetic pressure ratio decreases with 
$[z]$ and the warm model is Parker unstable. The dispersion relations show that 
the fastest growing mode has a wavelength of about 3 kpc, for both symmetric 
and antisymmetric perturbations, and the corresponding growth time scales are 
of about $3\times 10^7$ years. The structure of the final equilibrium stage is 
also derived, and we find that the midplane antisymmetric (MA) mode gathers 
more gas in the magnetic valleys. The resulting MA gas condensations have 
larger densities, and the column density enhancement is a factor of about 3 
larger than the value of the initial stage. The unstable wavelengths and growth 
times for the multi-component disk model are substantially larger than those 
of a thin disk model, and some of the implications of these results are 
discussed. 
\end{abstract}


\section{INTRODUCTION} 

In a series of seminal papers, Parker (1966, 1967, 1969) discussed the 
stability of a magnetized interstellar system with cosmic rays, and immersed 
in an external gravitational field. To build an equilibrium state in a 
plane-parallel density distribution, he assumed that i) the initial magnetic 
field is parallel to the galactic plane, ii) the gravitational acceleration 
is constant, iii) and the vertical pressure distributions for the gas, cosmic 
rays and magnetic field are simply described by an exponential function with 
the same scale height. Using a normal mode analysis he found that such a system
is unstable if the adiabatic index of the gas is below a certain critical 
value.  When the perturbation wavevectors are confined to the two-dimensional
(2D) plane defined by the directions of the initial magnetic and gravitational 
fields, the critical value is defined by $\gamma_{{\rm cr},u} = (1+\alpha+
\beta)^2/(1+ 0.5\alpha+ \beta)$, where $\alpha$ is the ratio of the 
magnetic-to-gas pressures, and $\beta$ is the ratio of cosmic-ray-to-gas 
pressures (Parker 1966). When the wavevectors are allowed to have all 
three-dimensional (3D) components, the critical adiabatic index becomes 
$\gamma_{{\rm cr},m} =1+\alpha+\beta$ (Parker 1967).

Later, these (2D and 3D) types of perturbations were classified as the {\it 
undular} (2D) and {\it interchange} (Hughes \& Cattaneo 1987) or {\it mixed} 
(3D) modes, (Matsumoto \etal 1993). Using the ``energy principle'' method, 
Lachi\`{e}ze-Rey \etal (1980) also found a generalized form for the critical 
adiabatic index (eq.~[9] in their paper), which is basically the same one of 
the mixed mode. The critical adiabatic index for this mixed mode is in general 
smaller, and more restrictive, than that of the undular mode. Given that 
cooling times in the diffuse interstellar medium (ISM) are shorter than the 
timescales for the instability, Parker used the isothermal value for the 
adiabatic index, $\gamma = 1$, and concluded that the equilibrium state of the
general ISM is then unstable.

The undular instability, which promotes the formation of high-density 
structures, is eventually stabilized by the tension of the distorted field 
lines. Using mass invariance along flux tubes, Mouschovias (1974) obtained the 
2D final equilibrium state of the original one-dimensional Parker model. Even 
when this new 2D equilibrium stage is in turn unstable against 3D perturbations 
(see Ass\'{e}o \etal 1978; Ass\'{e}o \etal 1980), the undulation pattern along 
the initial field lines persists in the nonlinear 3D evolution of the
instability (Kim \etal 1998). Therefore, the final 2D equilibrium state can be 
helpful in visualizing the resulting large-scale structure of the ISM.

The three original assumptions made by Parker described above are obvious 
idealizations, and some of them have been modified in subsequent studies. The 
first assumption, a well ordered field that is parallel to the galactic plane, 
is not really sustained by observations (except near the midplane). The 
interstellar ${\bf B}$-field has a bisymmetric spiral field configuration 
(see Heiles 1996; Indrani \& Deshpande 1998; Vallee 1998), and random 
components with cell sizes of the order of 50 pc (\eg Rand \& Kulkarni 1989). 
Also, the transition between the gaseous disk and the halo is very broad and 
has a complex structure with vertical field components (see Boulares \& Cox 
1990 for a discussion of the support provided by the tension of curved field 
lines). Thus, the plane-parallel field assumption is only valid as an average 
field configuration,  but it is very difficult to relax in both analytical and 
numerical treatments of the problem. As a variation to the simplest
plane-parallel field scheme, Hanawa, Matsumoto \& Shibata (1992) derived the 
unstable modes in a skewed magnetic field whose direction is still horizontal, 
but the field direction changes with distance from the midplane (\ie the $x$ and
$y$ components of the field vary with height, but $B_z$ is always equal to 
zero). For such a field configuration, the instability tends to form   
structures with the scales of giant molecular clouds.

The second assumption, a constant gravity, has been relaxed in more recent studies. The galactic gravitational field varies in a nearly linear fashion 
near the midplane (the linear approximation is excellent for $[z] \lesssim 150 
\ $ pc; Oort 1965; Bahcall 1984; Bienaym\'e, Robin \& Cr\'ez\'e 1987; Kuijken 
\& Gilmore 1989), and two different functions, linear and tanh ($z/H$), have 
been considered by different authors (Giz \& Shu 1993; Kim, Hong \& Ryu 1997; 
Kim \& Hong 1998). Since gravity is the driving force of the instability, a 
variation of the functional form for the acceleration has a direct impact on 
the properties of the unstable modes (\ie growth rates, length scales, and 
parity). For the constant gravity case, the weight of a gas parcel is the same 
regardless of its $z$-position, but the acceleration is discontinuous at 
$z=0$. Thus, the flow cannot move across the midplane and the only allowed 
modes are those with even parity. The resulting structures are then distributed 
symmetrically with respect to $z=0$, and are called midplane symmetric (MS). In 
the case of the other two functions (linear and tanh), the acceleration is 
continuous at the midplane and the weight of the gas increases with $[z]$. 
Thus, the odd parity solutions, or midplane antisymmetric (MA) modes, also 
appear, and the growth times are shorter than those of the uniform gravity case 
(\eg Giz \& Shu 1993; Kim \& Hong 1998). 

The third assumption, a disk with a single gas component, has not been modified
in any of the recent studies but it also has to be revised. The actual ISM
structure is very complex, and has several gas components ranging from the cold
molecular phase to a hot and highly ionized plasma (\eg Kalberla \& Kerp 1998).
Each component, in turn, has several sub-components with their own set of 
representative values for midplane densities and scale heights (the velocity 
dispersion of most components, however, seems to be equal to $\sim 9$ km 
s$^{-1}$: see Boulares \& Cox 1990). For instance, the atomic hydrogen phase 
could be divided into one cold H I component and two warm H I components (\eg 
Bloemen 1987; Boulares \& Cox 1990; McKee 1990; Spitzer 1990). To further 
complicate the situation, the vertical distributions for the magnetic field and 
cosmic rays do not seem to follow the stratification of the main gas 
components. Many of the system properties remain largely unknown and, 
depending on the assumed temperature and ${\bf B}$-field distributions, the 
resulting magnetohydrostatic (MHS) equilibrium configurations can be either 
stable or unstable to the Parker instability (\eg Bloemen 1987; Boulares \& Cox 
1990; Martos \& Cox 1994; Franco, Santill\'an \& Martos 1995; Kalberla \& Kerp 
1998). Thus, a quantitative stability analysis for this type of multi-component
gaseous disk is required.

In this paper we address this issue and investigate the 2D stability of an
extended, multi-component, magnetized disk with a ``realistic'' gravitational 
acceleration. We use the vertical equilibrium model for the warm magnetized
system that has been discussed by Martos (1993), Martos \& Cox (1994, 1998) and 
Santill\'an \etal (1999a). This equilibrium configuration is based on the 
observed distributions of: i) the vertical acceleration of the gravitational 
field in the solar neighborhood (Bienaym\'e, Robin \& Cr\'ez\'e 1987), and ii) 
the density distributions of the gaseous components (Boulares \& Cox 1990). 
Using the normal mode analysis, here we find that an isothermal extended disk 
is unstable with respect to the undular mode, and derive the resulting linear 
growth rates. We also derive the final 2D equilibrium state for both the MS and 
MA modes. The non-linear evolution is followed with the aid of 2D 
magnetohydrodynamic (MHD) numerical experiments and the results will be
presented in the accompanying paper by Santill\'an \etal (1999b, hereafter 
Paper II).

The plan of the present paper is as follows. In \S 2, we describe the initial 
MHS equilibrium state, and perform the normal mode analysis. The
dispersion relations for the unstable undular modes are then discussed. 
In \S 3, the nonlinear final equilibria of the undular modes are presented, 
and a summary and discussion of the results are given in \S 4.

\section{Normal Mode Analysis}

The linear stability analysis can be performed with either the ``energy 
principle'' method  (Bernstein \etal 1958) or with the usual ``normal mode'' 
analysis. The energy principle method provides the critical adiabatic index
in a relatively easy way (\eg Zweibel \& Kulsrud 1975; Ass\'eo \etal 1978; 
Ass\'eo \etal 1980; Lachi\`{e}ze-Rey \etal 1980), but does not allow to 
derive the resulting dispersion relations. Here we want to find 
the dispersion relations and, hence, derive them with the usual normal 
mode analysis.

\subsection{Isothermal Magnetohydrodynamic Equations}

The temperature distribution of the gaseous disk is largely unknown but, given
that the velocity dispersion of the main gas components are similar, we 
identify the velocity dispersion with the gas sound speed and define our model
as an isothermal disk stratification. Thus, we do not differentiate between the
thermal and kinetic pressures, and both are gathered together in a single 
pressure term with either a constant velocity dispersion or a constant 
effective temperature. The particular values for the resulting model velocity
dispersion and effective temperature are given below.

The dynamics of a magnetized isothermal plasma immersed in a gravitational 
field is described by the MHD equations,
\begin{equation}
\label{eq:continuity}
\frac{\partial \rho}{\partial t} + \nabla \cdot (\rho {\bf v}) = 0,
\end{equation}  
\begin{equation}
\label{eq:momentum}
\rho \left( \frac{\partial {\bf v}}{\partial t} 
           +{\bf v} \cdot \nabla {\bf v} \right)
= -\nabla \left( \rho a^2 + \frac{B^2}{8\pi} \right) 
  + \frac{1}{4\pi} {\bf B} \cdot \nabla {\bf B} + \rho {\bf g},
\end{equation}
\begin{equation}
\label{eq:induction}
\frac{\partial {\bf B}}{\partial t} = 
\nabla \times ({\bf v} \times {\bf B}),
\end{equation}
where $a$ (= constant) is the isothermal sound speed, and the rest of the 
symbols have their usual meanings. We use a Cartesian coordinate system 
($x,y,z$), whose axes are defined parallel to the radial, azimuthal, and 
vertical directions, respectively. We perform the analysis in the $y-z$ plane
and assume that the gravitational acceleration has only a vertical component, 
${\bf g} = [0,0,- g(z)]$.  

\subsection{Initial Equilibrium Configuration}

After Parker (1966) introduced the simplified exponential equilibrium model for
the gaseous disk, several authors have built more complex and realistic 
approximations for the actual ISM structure (\eg Badhwar \& Stephens 1977;
Bloemen 1987; Boulares \& Cox 1990; Kalberla \& Kerp 1998). These models are 
based on the observed stratifications for the gas, cosmic rays, and magnetic 
and gravitational fields in the solar neighborhood. Here we use, as the 
initial equilibrium state in our stability analysis, the MHS equilibrium 
configuration discussed originally in Martos (1993). This initial model uses 
the vertical distributions for the density and gravitational acceleration 
described by Boulares \& Cox (1990) and Bienaym\'e, Robin \& Cr\'ez\'e (1987), 
respectively. 

The density stratification is
\begin{eqnarray}
\label{eq:n}
n_0(z) &=& 0.6 \exp \left[ - \frac{z^2}{2(70 \mbox{pc})^2} \right]
       + 0.3 \exp \left[ - \frac{z^2}{2(135\mbox{pc})^2} \right]
       + 0.07\exp \left[ - \frac{z^2}{2(135\mbox{pc})^2} \right] \nonumber \\
     &+& 0.1 \exp \left[ - \frac{|z|}{400\mbox{pc}} \right] + 0.03 \exp 
       \left[ - \frac{|z|}{900\mbox{pc}}\right]~~~{\rm cm}^{-3}.
\end{eqnarray}
The midplane value is $n_0(0)\simeq 1.1 \ {\rm cm}^{-3}$ and, for a plasma with 
10\% He, the corresponding mass gas density stratification is $\rho_0(z)=1.27\ 
m_{\rm H} n_0(z)$, where $m_{\rm H}$ is the mass of a hydrogen atom. Figure~1 
shows the distribution of each gas component (representing the contributions of 
H$_2$, cold H I, warm H I in clouds, warm intercloud H I, and warm diffuse H II). 
The molecular and cold atomic phases are the dominant ISM mass components 
near the midplane, whereas the warm intercloud H I and warm diffuse H II are 
the most important gas layers beyond $z\sim 300$ pc. The extended H II component
was originally detected in absorption against the Galactic synchrotron 
background (Hoyle \& Ellis 1963), and later it was reported in hydrogen 
recombination emission (Reynolds 1989). This ionized gas is a major component
of the ISM, which has been usually ignored in previous modeling, and its 
surface density is about a third of that of the H I component at the solar 
neighborhood. The power requirements to ionize this layer are comparable to 
that available from supernovae. 

Because of the inclusion of the extended components, with scale heights larger 
than 300 pc, our model is referred to as the {\it thick} gaseous disk model. 
The resulting effective scale height is defined by the total gas column density 
as 
\begin{equation}
\label{eq:Heff}
H_{\rm eff} = \frac{1}{n_0(0)} \int_{0}^{\infty} n_0(z) dz \simeq 166 \ \ {\rm pc}.  
\end{equation}
In the case of the gravitational field, the observationally derived 
acceleration at the solar neighborhood can be fitted by (Martos 1993) 
\begin{equation}
\label{eq:gravity}
g(z) = 8 \times 10^{-9}
       \left[1 - 0.52 \exp\left(-\frac{|z|}{325 \mbox{pc}} \right)
               - 0.48 \exp\left(-\frac{|z|}{900 \mbox{pc}} \right) 
       \right]~~~{\rm cm~s}^{-2}.
\end{equation}
This gravitational acceleration is similar to the one derived by Kuijken \&
Gilmore (1989), and requires less local dark matter content than the ones derived by Oort (1965) and Bahcall (1984).

Given these two basic building blocks, $\rho_0(z)$ and $g(z)$, the initial 
equilibrium configuration is constructed by assuming that the gas is isothermal
and the initial magnetic field is parallel to the galactic plane. The effects
of cosmic rays are not explicitly included here because the results may depend 
on the assumptions made. For instance, in contrast to the effects of the 
isotropic cosmic ray pressure considered by Parker (1966), Nelson (1985) showed 
that an anisotropic cosmic ray pressure may tend to stabilize the gas layer.
For simplicity, then, we gather the non-thermal pressures into a single term 
represented by the magnetic pressure (\ie we assume that the sum of the cosmic 
ray and magnetic pressures is contained in the magnetic term). Then, the MHS 
equilibrium for the gas-field-gravity system is given by
\begin{equation}
\label{eq:iMHS}
\frac{d}{dz}P_0(z) = 
\frac{d}{dz} \left[\rho_0(z) a^2 + \frac{B_0^2(z)}{8\pi} \right] = 
- \rho_0(z) g(z),  
\end{equation}
where $P_0(z)$ is the total pressure of the system (thermal plus magnetic). This
equation defines the stratification of the magnetic field. For completeness, 
because our system is finite, we set the additional boundary condition $P_0 
(z=10 {\rm kpc})=0$, and the system pressure is computed with the integral
\begin{equation}
P_0(z) = \int_{z}^{10{\rm kpc}} \rho_0(z) g(z) dz.
\end{equation}
Given the total pressure and the strength of the magnetic field (including both
ordered and random components) at midplane, $P_0(0) \sim 3\times 10^{-12}$ dyne 
cm$^{-2}$ and $B_0(0) \simeq 5\mu$G (Boulares \& Cox 1990; Heiles 1996), the 
resulting isothermal sound speed (from $P_0(0) = 1.27\ m_{\rm H} n_0(0) a^2 + 
B_0^2(0)/8\pi$, with $n_0(0) = 1.1$ cm$^{-3}$) is $a=8.4$ km s$^{-1}$. Thus, 
the sound speed value is very similar to the observed velocity dispersion of 
the main gas components (within 5 to 9 km s$^{-1}$; Boulares \& Cox 1990), and 
the corresponding effective disk temperature is $T_{\rm eff}= 10900$ K. This 
is called the ``warm'' magnetic disk model and its properties are discussed by 
Martos (1993), Martos \& Cox (1994, 1998) and Santill\'an \etal (1999a). 
Figure~2 shows the distributions of the thermal, magnetic, and total pressures 
as functions of distance from the galactic plane. The maximum of the magnetic 
pressure is not centered at $z=0$ because the field stratification is derived 
from MHS equilibrium. This warm magnetic disk model is Parker unstable because 
the gas is almost entirely supported by the magnetic field above $z \simeq 
200$~pc.

There is high-latitude H I gas with velocity dispersions of 35 km s$^{-1}$ 
(Kulkarni \& Fich 1985), and halo gas with up to 60 km s$^{-1}$ (Kalberla \etal 
1998). The inclusion of these additional gas components with different velocity 
dispersions in our analysis is beyond the scope of the present paper but, as
sketched in \S 4, we will address this issue in a future study. 

\subsection{Linearized Perturbation Equations}

We limit the present discussion to perturbations in the $y-z$ plane (\ie in the
plane defined by the directions of the initial magnetic and gravitational 
fields). Due to this limitation, only the undular modes are allowed and we 
follow the procedure described by Kim, Hong \& Ryu (1997) and Kim \& Hong 
(1998) to derive the properties of the instability. Given that the velocities 
in the initial model are equal to zero, we denote by ${\bf v}$, $\delta \rho$, 
$\delta {\bf B}$ the infinitesimal perturbations in velocity, density, and 
magnetic field, respectively. The perturbed state is then described by 
\begin{equation}
\label{eq:pstate}
{\bf v};~~~~~\rho = \rho_0 + \delta \rho;~~~~~
{\bf B} = B_0\hat{e}_y + \delta {\bf B}.
\end{equation}
Inserting these perturbed variables in equations 
(\ref{eq:continuity}-\ref{eq:induction}), and keeping only the first-order 
terms for the perturbations, the linearized perturbation equations become
\begin{equation}
\label{eq:rho}
\frac{\partial}{\partial t} \delta\rho + v_z\frac{d\rho_0}{dz} 
+ \rho_0 \left( \frac{\partial v_y}{\partial y} 
               +\frac{\partial v_z}{\partial z} \right) =0,
\end{equation}
\begin{equation}
\rho_0 \frac{\partial v_y}{\partial t} + 
\frac{\partial}{\partial y}(a^2 \delta\rho) 
- \frac{1}{4\pi}\frac{dB_0}{dz}\delta B_z = 0,
\end{equation}
\begin{equation}
\rho_0 \frac{\partial v_z}{\partial t}
+ \frac{\partial}{\partial z}(a^2\delta\rho) 
+ \frac{1}{4\pi}\frac{dB_0}{dz}\delta B_y
+ \frac{1}{4\pi}B_0\frac{\partial}{\partial z}\delta B_y
- \frac{1}{4\pi}B_0\frac{\partial}{\partial y}\delta B_z
+ g\delta\rho = 0,
\end{equation}
\begin{equation}
\frac{\partial}{\partial t} \delta B_y
+ B_0\frac{\partial v_z}{\partial z}
+ \frac{dB_0}{dz}v_z = 0,
\end{equation}
\begin{equation}
\label{eq:Bz}
\frac{\partial}{\partial t}\delta B_z 
- B_0\frac{\partial v_z}{\partial y} = 0.
\end{equation}
The coefficients of equations (\ref{eq:rho}-\ref{eq:Bz}) do not depend 
explicitly on $y$ and $t$, and the perturbations can be Fourier-decomposed with 
respect to these variables
\begin{equation}
\left[
      \begin{array}{c}
             \delta\rho(y,z;t) \\
             v_y(y,z;t)        \\
             v_z(y,z;t)        \\
             \delta B_y(y,z;t) \\
             \delta B_z(y,z;t) 
      \end{array} 
\right]
=
\left[
      \begin{array}{c}
             \delta\rho(z) \\
             v_y(z)        \\
             v_z(z)        \\
             \delta B_y(z) \\
             \delta B_z(z) 
      \end{array} 
\right]
\exp(i\omega t - ik_y y),
\end{equation}
where $i \omega$ is the growth rate and $k_y$ is the wavenumber along the
$y$-direction. Inserting these decomposed forms into the perturbation equations 
(\ref{eq:rho}-\ref{eq:Bz}), and combining them we obtain the reduced 
equation,
\begin{equation}
\label{eq:vz}
f\frac{d^2v_z}{dz^2} + \frac{df}{dz}\frac{dv_z}{dz} + hv_z = 0,
\end{equation}
where the functions $f$ and $h$ are defined by
\begin{equation}
f = 2 (\omega^2  - k_y^2 a^2) \frac{B_0^2}{8\pi} + \omega^2 \rho_0 a^2,
\end{equation}
\begin{equation}
h =  (\omega^2 - k_y^2a^2)
     \left(\omega^2 \rho_0 - 2k_y^2\frac{B_0^2}{8\pi} \right)
    - \omega^2 \rho_0 \frac{dg}{dz}  + k_y^2 g \frac{d}{dz} \left(\frac{B_0^2}{8\pi}\right).
\end{equation}
The factor $dg/dz$ appearing in the second term of $h$ is introduced by taking 
the derivative with respect to $z$ on both sides of the MHS equation 
(\ref{eq:iMHS}), and then making the appropriate substitutions. This results in
a more compact form for the $h$ function (and we do not need to calculate 
numerically a second-order derivative term). With the transformation $\Psi=v_z 
f^{1/2}$, equation~(\ref{eq:vz}) can be rearranged to 
\begin{equation}
\label{eq:master}
\Psi^{\prime\prime} 
+ \left[\frac{1}{4}\left(\frac{f^\prime}{f}\right)^2
-\frac{1}{2}\left(\frac{f^{\prime\prime}}{f}\right)
+\frac{h}{f} \right] \Psi = 0,
\end{equation}
where the prime superscript ($^\prime$) denotes the derivative with respect to
$z$. Given the complicated functional forms for $\rho_0(z)$, $g(z)$, and $B_0(z)$,
one cannot perform further simplications of equation~(\ref{eq:master}). 

The required boundary conditions (BCs) are: $\Psi=0$ at an upper boundary $z=
z_{\rm node}$, and $\Psi=0$ or $d\Psi/dz=0$ at the midplane, $z=0$. The first 
condition at the midplane, $\Psi=0$ at $z=0$, generates the even parity MS 
solutions, whereas the second one corresponds to the odd parity MA solutions 
(\eg Horiuchi \etal 1988; Giz \& Shu 1993). 

\subsection{Dispersion Relations}

The dispersion relations are found with the method described in the Appendix
of Kim \etal (1997).  The method is a numerical procedure to find, for a given
wavenumber, an eigenvalue ($i \omega$) which satisfies the imposed BCs. Our 
equilibrium configuration, as stated in \S 2.2, turns out to be Parker 
unstable, and we find eigenvalues $i \omega$ that are real and positive. The 
resulting dispersion relations are shown in Figure~3 for five cases whose 
upper boundaries are placed at $z$-locations ranging from $z = 9H_{\rm eff}$ to 
$z = 30H_{\rm eff}$. The growth rates, wavenumbers, and nodal points in Figure 
3 are normalized as:
\begin{equation}
\Omega = i \omega \frac{H_{\rm eff}}{a},~~~\nu_y = k_y H_{\rm eff},~~~
\zeta_{\rm node} = \frac{z_{\rm node}}{H_{\rm eff}}.
\end{equation}
Since gravity has small values near the midplane, the dispersion relations are
not sensitive to the midplane boundary conditions. Thus, the solutions are 
degenerate with respect to parity, and the growth rates are nearly the same for 
both the MS and MA modes. The plotted dispersion relations are for the 
principal $z$-modes, whose MS nodal points are located at midplane and 
$\zeta_{\rm node}$. The nodal points of the MA modes are located at 
$-\zeta_{\rm node}$ and $\zeta_{\rm node}$. For the lower nodal point in 
Figure~3, $\zeta_{\rm node}=9$ ($z_{\rm node}= 1.5$ kpc), the fastest growth 
time is about $6.2 \times 10^7$ years and its wavelength is 3.11 kpc. For the
upper point, $\zeta_{\rm node}=30$ ($z_{\rm node}$=5 kpc), the corresponding 
values change to $3.4 \times 10^7$ years and 3.43 kpc, respectively. From the
figure, it is clear that the maximum growth rate above $z \ge 3$ kpc is less 
sensitive to the position of the nodal point. This is because the gravitational
acceleration (see eq.~[\ref{eq:gravity}]) already reaches its maximum value, 
$8 \times 10^{-9}$ cm s$^{-2}$ at about $\sim 3$ kpc. Therefore, as the nodal
point goes to positions higher than $\zeta_{\rm node}=30$, the growth time 
converges to $\sim 3 \times 10^7$ years, which can be regarded as the minimum 
growth time of the Parker instability in the thick gaseous disk (obviously, at
these heigths the gravitational force also has a non-neglegible radial 
component, and we are near the limit of validity of our 2D analysis). In the 
following section we address the structure of the final equilibrium state.

%The evolution of the linear and non-linear regimes has been performed with 
%detailed MHD numerical experiments in 2-D, and the results are presented in 
%Paper II. The initial linear phase of these numerical experiments follow the
%rates and the wavelengths derived in the present linear analysis, and 

\section{Two-dimensional Equilibria of the Undular Instability}

\subsection{Magnetohydrostatic Equations}

The MHS equations are obtained by setting ${\bf v}=0$ and dropping the 
time-derivative terms in the MHD equations
(\ref{eq:continuity}-\ref{eq:induction}). Hence, the continuity and induction 
equations are of no use in this case. Due to this reason, the number of 
unknowns ($\rho$, $B_y$, and $B_z$) is larger than the number of equations 
(the $y$ and $z$ components of the momentum equation), and one requires an
additional expression. Closure is granted with flux freezing, which 
results in conservation of the mass-to-flux ratio in a flux tube.

The details for the derivation of the final equilibrium states are given in 
Mouschovias (1974), and are summarized in Spitzer (1978). Given the magnetic 
vector potential ${\bf A}=\hat{e}_x A(y,z)$
\begin{equation}
{\bf B} = \nabla \times {\bf A},
\end{equation}
and the gravitational potential
\begin{equation}
\psi = \int_{0}^{z} g(z) dz,
\end{equation}
the final magnetic equilibrium is given by
\begin{equation}
\label{eq:Poisson}
\nabla^2 A = -4\pi \frac{dq}{dA} 
                   \exp \left( -\frac{\psi}{a^2} \right).
\end{equation}
The function $q \equiv \rho a^2 \exp(\psi/a^2)$ is a constant along a line of
force and is given by 
\begin{equation}
\label{eq:q}
q(A) = \frac{a^2}{2}\frac{dm}{dA}
       \left\{ \int_{0}^{{\lambda_y}/2}dy
                    \frac{\partial z(y,A)}{\partial A}
       \exp \left[ -\frac{\psi(y,A)}{a^2} \right] \right\}^{-1},
\end{equation}
where $\lambda_y$ is the perturbation wavelength along the initial magnetic 
field, and $dm/dA$ is the mass-to-flux ratio. As stated above, for flux 
freezing conditions the mass between two field lines is conserved and the 
mass-to-flux ratio remains constant during the evolution (\ie is a constant of 
motion). Then, this ratio is determined from the initial equilibrium 
configuration 
\begin{equation}
\label{eq:dmda}
\frac{dm}{dA} = \lambda_y \frac{\rho_0(A)}{B_0(A)},
\end{equation}
where $\rho_0(A)$ and $B_0(A)$ represent the initial distributions of the 
density and magnetic field as functions of $A$, respectively.

\subsection{Final Equilibrium States}

Now, after setting the mass-to-flux ratio, one can solve equations 
(\ref{eq:Poisson}) and (\ref{eq:q}) simultaneously. Following the detailed 
procedure described in Appendix C of Mouschovias (1974), we solve these 
equations by iteration. In contrast to the original work of Mouschovias, who 
used a constant gravity with a discontinuity at midplane, we use a smooth
and continuous gravity function (eq.~[\ref{eq:gravity}]). The discontinuity 
prevents midplane gas crossings, and he found the final equilibria of the MS 
modes only. We do not have such a discontinuity and are able to derive the 
final equilibria of both the MS and MA modes.  

The initial equilibrium distributions for the density and field lines are 
plotted in Figure~4a. Colors are mapped from red to violet as the density 
decreases. The white lines represent the ${\bf B}$-field lines, and they 
are chosen in such a way that the magnetic flux between two consecutive 
lines is the same. The length scales are normalized with the effective 
scale height, $H_{\rm eff}$. First, to derive the final state of the MS mode, 
we added a MS perturbation to the magnetic vector potential in the initial 
equilibrium state,
\begin{equation}
\delta A(y,z) = -A_0(z) C \cos(\frac{2\pi y}{\lambda_y}) 
                            \sin(\frac{2\pi z}{\lambda_z}),
\end{equation}
where $C=0.01$ is the amplitude of the perturbation, and $\lambda_z = 2 
z_{\rm node}$ (the MS mode has zero amplitude at $z=0$ and $z=z_{\rm node}$,
so $z_{\rm node}$ corresponds to a half of the wavelength value of the 
principal mode along the $z$-axis). Figure~3 shows that when the first nodal 
point from midplane is $9H_{\rm eff}$, the most unstable horizontal wavelength 
is $18H_{\rm eff}$. Thus, we use the pair of unstable wavelengths $(\lambda_y,
\lambda_z)=(18H_{\rm eff},18H_{\rm eff})$, and get the final equilibrium state 
displayed in Figure 4b. The actual computational domain for this symmetric 
case is $0 \le y \le 9H_{\rm eff}$ and $0 \le z \le 9H_{\rm eff}$, but for 
visualization purposes we extend the domain eight times in the figure.  
The condensations formed in the magnetic valleys and voids in the arches are 
clearly seen in the figure. Due to the condition imposed at midplane, 
$\delta A =0$, the field line at $z=0$ is not deformed at all. 

For the MA case, we perturb the initial state with the perturbation,
\begin{equation}
\delta A(y,z) = -A_0(z) C \cos(\frac{2\pi y}{\lambda_y}) 
                            \cos(\frac{2\pi z}{\lambda_z}),
\end{equation}
where we now use $\lambda_z = 4 z_{\rm node}$ (in contrast to the MS mode, 
the MA mode has maximum amplitude at $z=0$ and then requires twice the 
wavelength value along the $z$-axis). As stated before, the dispersion 
relations shown are not sensitive to the midplane boundary conditions, and the 
most unstable horizontal wavelength is the same for both the MS and MA modes.
Using the same nodal point as before, $z_{\rm node}=9H_{\rm eff}$, we set the
pair of unstable wavelengths to $(\lambda_y,\lambda_z)=(18H_{\rm eff},36H_{\rm 
eff})$, and the final state of the MA mode is plotted in Figure~4c. The 
computational domain is now $0\le y\le 9H_{\rm eff}$ and $-9 H_{\rm eff}
\le z \le 9H_{\rm eff}$, 
and covers the upper and lower hemispheres. For a better visual 
impression we extend it by a factor of four in Figure~4c. As before, one 
can also see condensations and voids in the figure, but their positions are now
alternated between the upper and lower hemispheres. Hence the distance between 
successive condensations is a half of the horizontal wavelength. Also, the 
${\bf B}$-field line at midplane is now undulated, with locations above and 
below $z=0$, as is characteristic of the MA mode.     

%A set of MHD simulations with different velocity perturbations, and in more 
%extended computational domains including $(54H_{\rm eff},18H_{\rm eff})$, are 
%presented in Paper II. The numerical results for the most unstable wavelength
%with random velocity perturbations are similar to Fig.~4c, and indicate that 
%the MA mode is preferred over the MS mode.

The density enhancements produced by the gas that has been sliding into the
magnetic valleys can be obtained from the column density of the final state. At
any given location $y$, the final column density is  
\begin{equation}
\label{eq:colden}
N_f(y) = \int_{z(y,A_i[z=0])}^{9H_{\rm eff}} \rho_f(y,z) dz,
\end{equation}
where the subscripts $i$ and $f$ denote the initial and final states, 
respectively. The lower limit of the integral corresponds to the final
$z$-coordinate of the magnetic field line that was initially located at 
midplane, and is labeled with $A_i(z=0)$. Thus, the lower limit is exactly 
equal to zero for the MS modes, but it is different from zero for the MA modes. 
In Figure~5 we plot the column density distributions for both modes (normalized
to the initial column density, $N_i=\int_0^{9H_{\rm eff}}\rho_i(z)dz$). The 
figure reveals that the MA modes drive more gas into the magnetic valleys than 
the MS cases. This is because the MA perturbations can gather more mass in the
condensations by bending the midplane.

Another interesting quantity is the ratio of the magnetic-to-gas pressures,
$\alpha=B^2/(8\pi a^2 \rho)$. The value of this ratio varies with time and $z$-location, and Figure~6 shows the corresponding distributions at the initial 
and final equilibrium states. The initial state is plotted as a solid line, and 
is labeled as $\alpha_i (z)$. The distribution for the final MS state is 
plotted with dashed lines at two different $y$-positions: the distribution at
$y=0$, corresponding to the center of the condensation (the magnetic valley),
is labeled as $\alpha_f (0,z)$, and the distribution at $y=9H_{\rm eff}$ (the
central part of the magnetic arch; see Fig. 4b) is labeled as $\alpha_f (9,z)$. 
Finally, the distribution for the final MA state at the position $y=0$ is shown 
with a dotted line, and is also labeled $\alpha_f (0,z)$ (the lower part of the
disk contains the central part of a condensation, and the upper part of the 
disk has the maximum of the magnetic arches; see Fig. 4c). The 
$\alpha$-distribution of the initial state shows that our disk model is 
mainly supported by gas pressure near the midplane, and by magnetic pressure at 
high latitudes (see also Fig.~2). The distribution, however, is completely 
modified at the final equilibrium stages. The gas pressure increases at the 
condensations and the resulting pressure ratios, for both the MS and MA modes, 
become smaller than the initial $\alpha$ values. In contrast, at the voids, 
where the magnetic energy becomes dominant, $\alpha$ reaches values as high as 
$\simeq 10^4$. This is because the gas is efficiently drained down from the 
magnetic arches, as already pointed out by Mouschovias (1974) for the case of 
a thin gaseous disk in a uniform gravity.

As a final comment of this section, we add that the galactic system seems to
prefer the lower energy state of the MA mode. This is not apparent from the 
dispersion relations, which are degenerate for the MS and MA modes, but it 
appears in detailed numerical MHD experiments performed with the same 
thick disk model considered in this study. The results of these numerical 
simulations will be reported in a separate paper (Santill\'an \etal 1999b), and 
here we only mention the relevant result. The runs are started with the initial 
equilibrium state described \S 2.2 and, as expected, the early linear phase of 
the experiments follows the rates and wavelengths derived in the present linear 
analysis for any of the two parity modes. We also performed several experiments 
with random velocity perturbations, without any preferred parity or wavelength. 
These random perturbation experiments eventually evolve into the MA 
configuration, similar to the one shown Figure~4c, indicating that the MA mode 
is preferred over the MS mode. 

%By bending the midplane, the MA mode drains down more gas into lower positions 
%and releases more gravitational energy than the MS mode.  

\section{Summary and Discussions}

Here we have presented the linear perturbation analysis of a magnetized and 
warm thick disk. The dispersion relations for the undular mode of the Parker 
instability are derived, along with the resulting final equilibrium states. 
The initial disk parameters are taken from the observed distributions at the 
solar circle, and we assume that the gas is isothermal and that the initial 
field lines are parallel to the disk. Given the complexities inherent to trying
to model the dynamical effects of cosmic rays, its pressure is not explicitly 
included in here. The resulting 
multi-component gaseous disk model, then, has a thermal-to-magnetic pressure 
ratio that decreases with $z$-location, and is Parker unstable. The properties 
of the unstable modes for five different nodal points 
are given in Figure 3. These nodal points 
correspond with the assumed extension of the disk above midplane (for $H_{\rm 
eff}\simeq 166$ pc, the five cases in Figure 3 represent a disk extending up 
to 1.5, 2, 3, 4 and 5 kpc, respectively). 

The value of the critical wavelength for the instability depends on the 
location of the chosen nodal point and, for disks extending between 5 and 1.5 
kpc above midplane, it increases from 1.5 to 1.8 kpc, respectively. The 
wavelength of the fastest growing mode, however, is almost independent of the
assumed nodal point, and is about 3 kpc for all the cases considered. The 
corresponding growth time scales are slightly more sensitive to the nodal  
point and, for disks extending between 1.5 and 5 kpc above midplane, the 
time scales decrease from 6.2 to 3.4$\times 10^7$ year, respectively. The
minimum growth time then converges to $\sim 3 \times 10^7$ year as the nodal
point tends to large $z$-values. Thus,
for average conditions in the solar neighborhood, the linear analysis in 2D 
indicates that the preferred wavelength is about 3 kpc, and that the gas 
condensations are formed in time scales of the order of about $\sim 3\times 
10^7$ year. The wavelength values are larger, by a factor of about 8, than 
those derived for the thin disk cases, but the corresponding time scales are 
larger by only a factor of about 2 (Kim \& Hong 1998). These are substantial
differences, and indicate that the multi-component structure of the disk play 
an important role in the large-scale stability and evolution of the ISM. 

The densities for the final equilibrium stages, on the other hand, are larger
for the MA modes. The resulting MA column densities at the condensations are 
increased by a factor of about 3 with respect to the value of initial
equilibrium stage. Now, the spiral density wave can trigger the Parker 
instability in the model considered here (Martos \& Cox 1994), and the contrast
obtained is similar to the expected density contrast between arm-interarm 
regions for strong waves (Elmegreen 1991). For comparison, the final-to-initial 
column density ratio of the fastest growing mode in a thin disk model is of 
about 1.2 only (Mouschovias 1974). Thus, the gas from the extended gas layers
participating in the instability contribute with a fraction of about 2/3 to 
the total mass gathered in the condensations. 

The role of self-gravity and differential rotation of the Galaxy are not 
included in the present study. Self-gravity may not be important at the early
linear stages of the instability (\eg Hanawa, Nakamura \& Nakano 1992), but it 
will lead to more compact and denser condensations at the non-linear phases. 
Galactic differential rotation, on the other hand, has an influence at several
stages of the Parker instability (\eg Shu 1974; Zweibel \& Kulsrud 1975; Balbus
\& Hawley 1991; Foglizzo \& Tagger 1994, 1995). For instance, if the radial 
differential force is strong enough, a transient shearing instability also appears, and the combined Parker-shearing instability could lead to angular
momentum transfer and dynamo action in disks. In the case of the 2-D undular
perturbations considered in this paper, the lateral motions of
the flows should be affected by the Coriolis force. If we, however, include the
ignored third dimension (the radial direction) in our analysis, the vertical 
motion of the mixed mode with a smaller wavelength along the radial direction 
dominates the system, and the effects of rotation are severely reduced during
the linear growth. Nonetheless, as stated by the referee, the stabilizing 
effects of rotation may be important at the final equilibrium stages. These are
important issues that require detailed three dimensional studies with differential rotation, and should be addressed in future studies.

If the assumptions of the present work are valid, the range of growth rate 
values are marginally consistent with those required for the formation of giant 
molecular clouds in our Galaxy (\eg Blitz \& Shu 1980). Also, the most unstable
wavelength in our model is somewhat larger than the corrugation distance 
derived by Alfaro \etal (1992) for the Carina Arm (2.4 kpc), but the 
condensations formed by the odd parity mode of the instability may well be 
associated with the origin of this observed structure. A more detailed study is 
required to properly address this issue, and important caveats should be borne 
in mind regarding the applicability of the present results. One is that of the
randomness of the Galactic magnetic field topology at the kpc length scale, not
included in the present modeling. Another one is the largely unknown 
temperature structure of the halo, and the filling factors of the different gas
components. Models built from the same density and gravity distributions, but 
in which the magnetic field distribution is prescribed from the Galactic 
synchrotron emission (\eg Martos \& Cox 1998), require thermal dominance at 
high $[z]$ and are therefore Parker stable. 

%As stated above for the
%Carina Arm, the length scales for the multi-component disk model seem to be 
%in better agreement with the characteristic corrugation distances observed 
%inside spiral arms in both our Galaxy and external spirals (\eg Quiroga \& 
%Schlosser 1977; Spicker \& Feitzinger 1986; Florido \etal 1991; Alfaro \etal 
%1992). 

This leads us to a final important question if the isothermal disk assumption 
represents a fair description of the actual gaseous disk in our Galaxy. Here we 
do not differentiate between the thermal and kinetic pressures, and both are 
gathered in a single isothermal term with sound speed similar to the velocity 
dispersion of the main components extending up to $\sim 1.5$ kpc from midplane 
(Boulares \& Cox 1990). Such an isothermal condition, then, can be considered 
as a reasonable approximation for the regions located between 1 to 1.5 kpc 
from the midplane. The existence of a few ``anomalous'' velocity components 
within 2 kpc (\eg Kulkarni \& Fich 1985; Reynolds 1985), and gas with a large 
velocity dispersion at $z$ of about 4 kpc (Kalberla \etal 1998), already hint 
that the effective sound speed should be increased somewhere within 1 and 2 
kpc. The details for such a variation are presently unknown, but we are 
currently investigating the effects of some reasonable velocity distributions. 
Obviously, the loss of magnetic support provides a stabilizing effect, but the 
present restrictions do not indicate that the instability can be completely 
suppressed. A detailed discussion of the range of velocity dispersion 
variations and the resulting unstable mode values will be presented elsewhere.

\acknowledgments
It is a big pleasure to thank Emilio Alfaro, Don Cox, Gene Parker, and Dongsu 
Ryu for many stimulating and informative discussions during the development of 
this project. We are grateful to Thierry Foglizzo, the referee, and Steve 
Shore, the editor, for several constructive comments. JF thanks the Korea 
Astronomy Observatory and Seoul National University for their warm hospitality. 
JF, MM and AS acknowledge partial support by DGAPA-UNAM grant IN130698, CONACyT 
grants 400354-5-4843E and 400354-5-0639PE, and by a R\&D CRAY Research grant. 
The work of JK was supported by the Office of the Prime Minister through Korea 
Astronomy Observatory grant 99-1-200-00, and he also acknowledges the warm 
hospitality of the Instituto de Astronom\'{\i}a-UNAM. The work of SSH was 
supported in part by a grant from the Korea Research Foundation made in the 
year 1997.

\clearpage

\centerline{\bf Figure Captions}

\figurenum{1}
\figcaption{
Vertical distribution of the ISM components. The gas is divided in three
components (molecular, neutral, and ionized), and the neutral component further
divided into three (cold, warm cloud, and warm intercloud) sub--components.
The number density for each component is plotted as a function of distance from 
midplane.
}

\figurenum{2}
\figcaption{
Pressures (gas, magnetic, and total) as a function of distance from midplane.
}

\figurenum{3}
\figcaption{
Dispersion relations of the undular instability in a magnetized multi-component
gaseous disk. Each curve is marked by the value of an upper nodal point,
$\zeta_{\rm node}$. The five nodal points shown, $\zeta_{\rm node}=$ 9, 12, 18, 
24, and 30, correspond to 1.5, 2, 3, 4, and 5 kpc, respectively. The ordinate 
corresponds to the square of the normalized growth rate, and the abscissa to 
the square of the normalized horizontal wavenumber. The normalization units are
the isothermal sound speed, 8.4 km s$^{-1}$, and the effective scale height, 
$H_{\rm eff}$=166 pc.
}

\figurenum{4}
\figcaption{
The initial and final states of a magnetized multi-component gaseous disk.
(a) Initial state, (b) final state for a midplane symmetric (MS) perturbation,
and (c) final state for a midplane antisymmetric (MA) perturbation.  Colors are
mapped from red to violet as the natural logarithmic value of density 
decreases, and white lines represent magnetic field lines. Nineteen field
lines are chosen in such a way that the magnetic flux between two consecutive
lines is the same. The uppermost and lowermost field lines lie exactly on the
upper and lower computational boundaries. The unit length is the effective 
scale height.
}

\figurenum{5}
\figcaption{
Column density for the final states as a function of the horizontal coordinate. 
It is given by the numerical integration of equation~(\ref{eq:colden}). The 
lower limit of the integral is the final $z$-coordinate of the field line which
is initially located at the midplane. The column density is normalized to its 
initial value, and the unit length is the effective scale height.
}

\figurenum{6}
\figcaption{
The ratio of the magnetic-to-gas pressures, $\alpha$, in the initial and final 
states. The solid line represents the initial $\alpha$ distribution. The two 
dashed lines give $\alpha$ values along the lines $y=0$ and $y=9$ in the 
final MS state, and the dotted line gives $\alpha$ along $y=0$ in the 
final MA state. The unit length is the effective scale height.
}

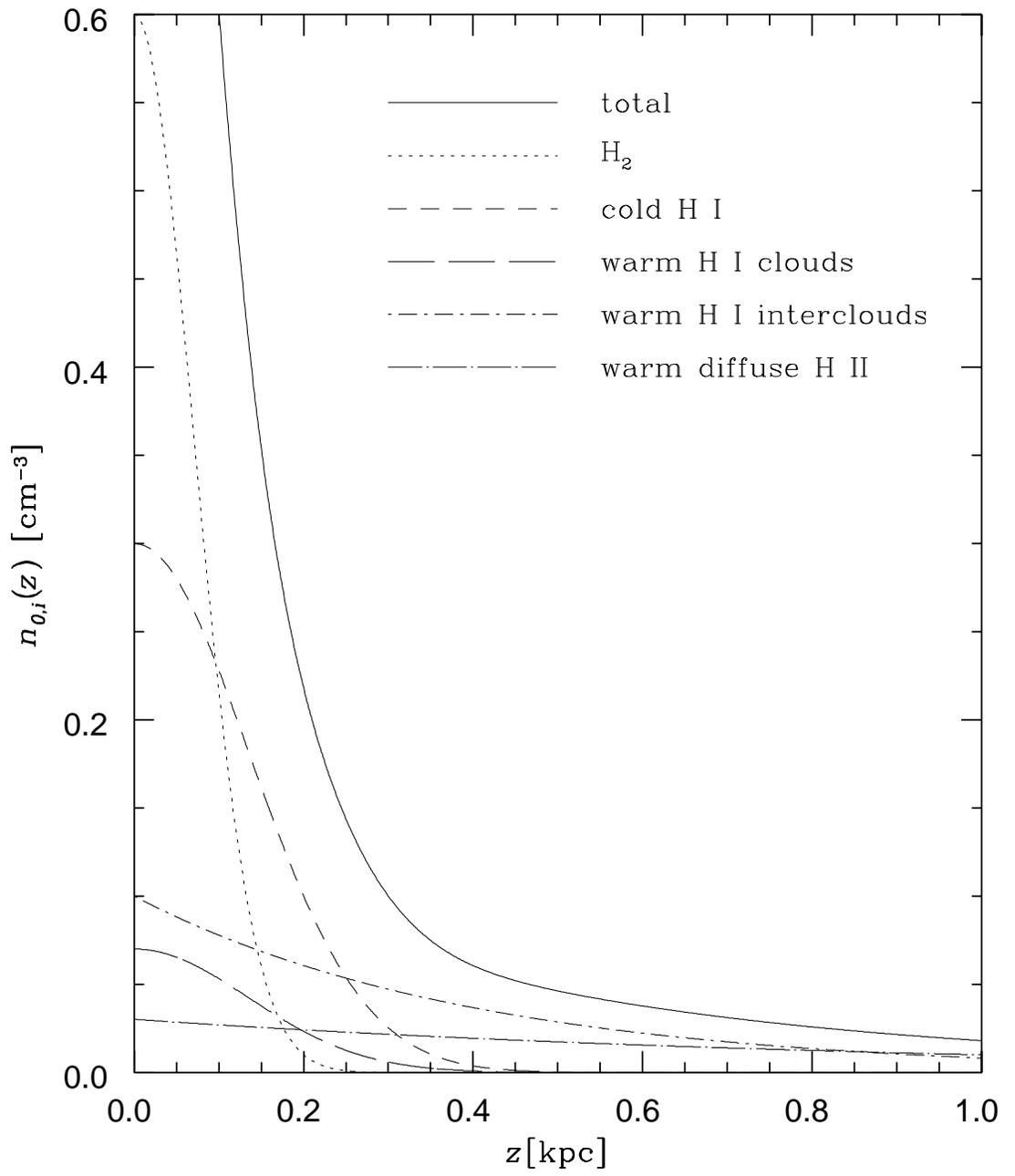
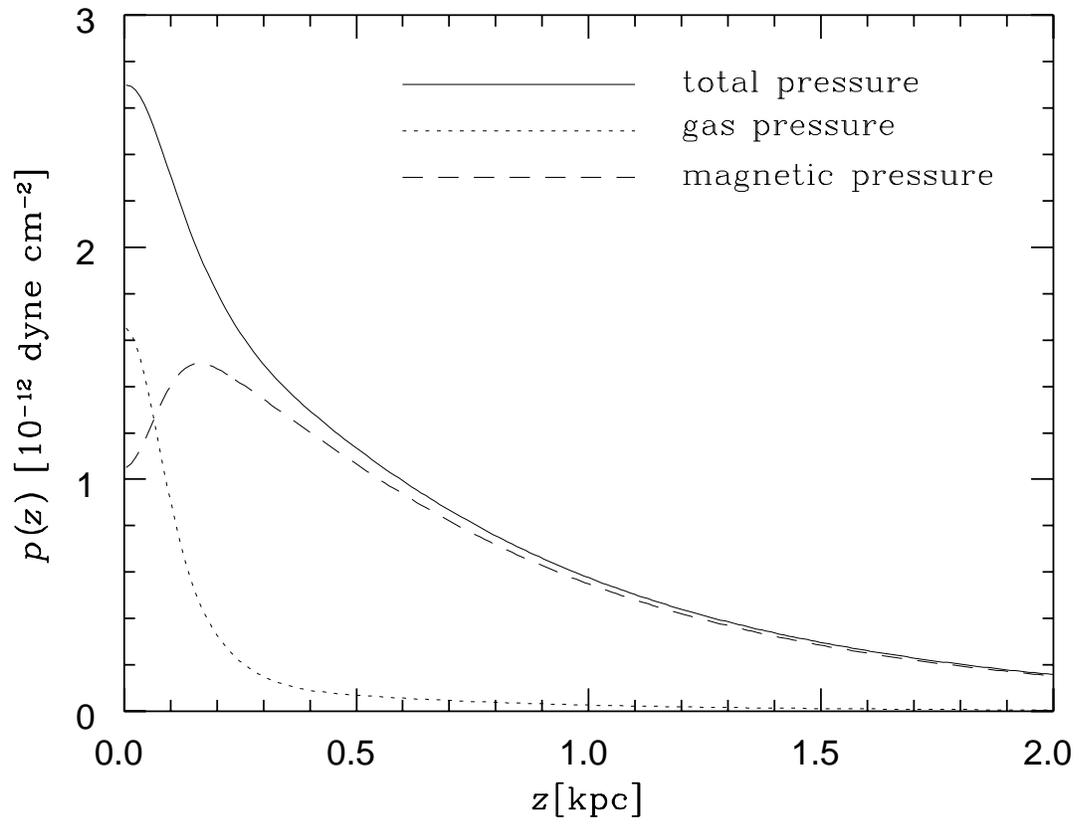
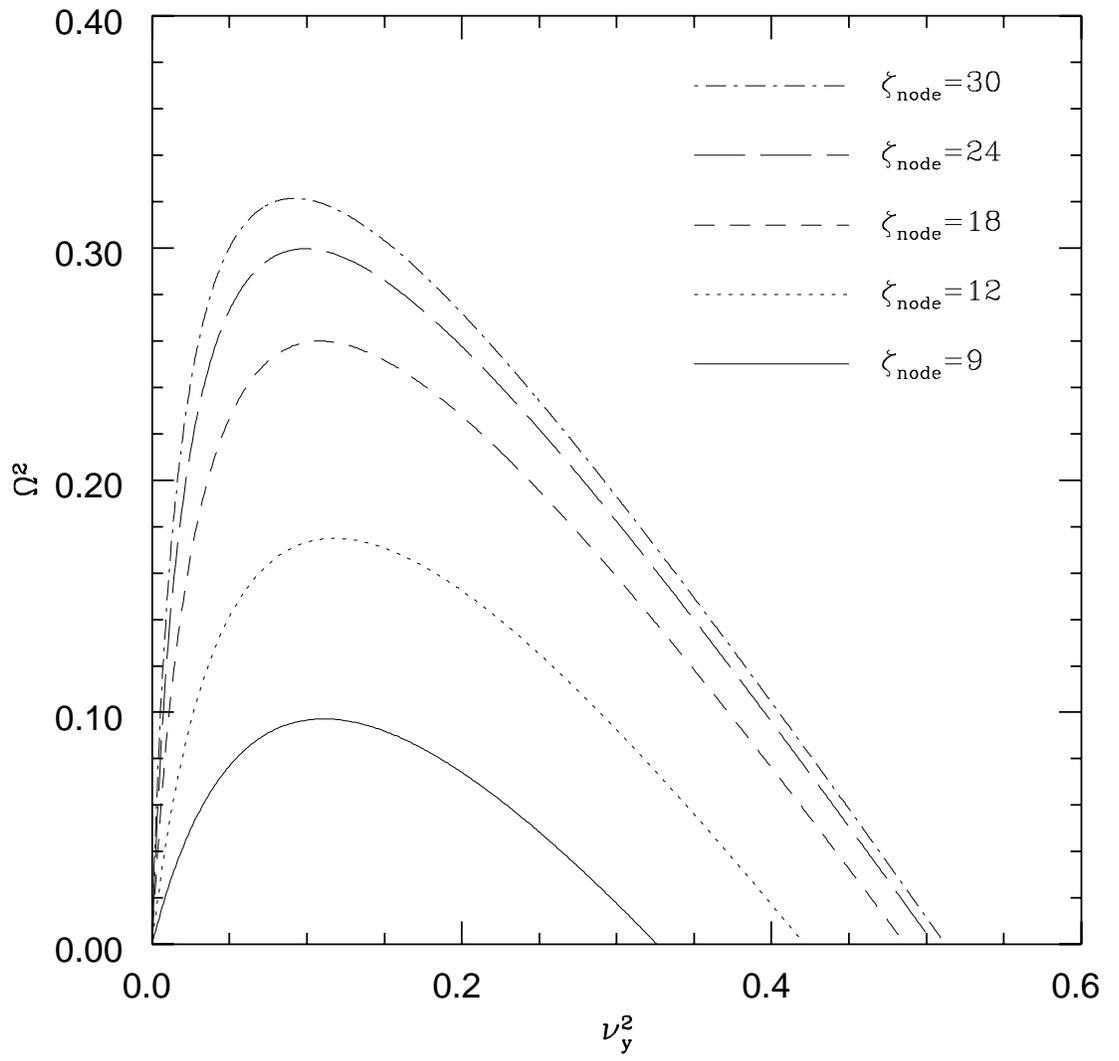
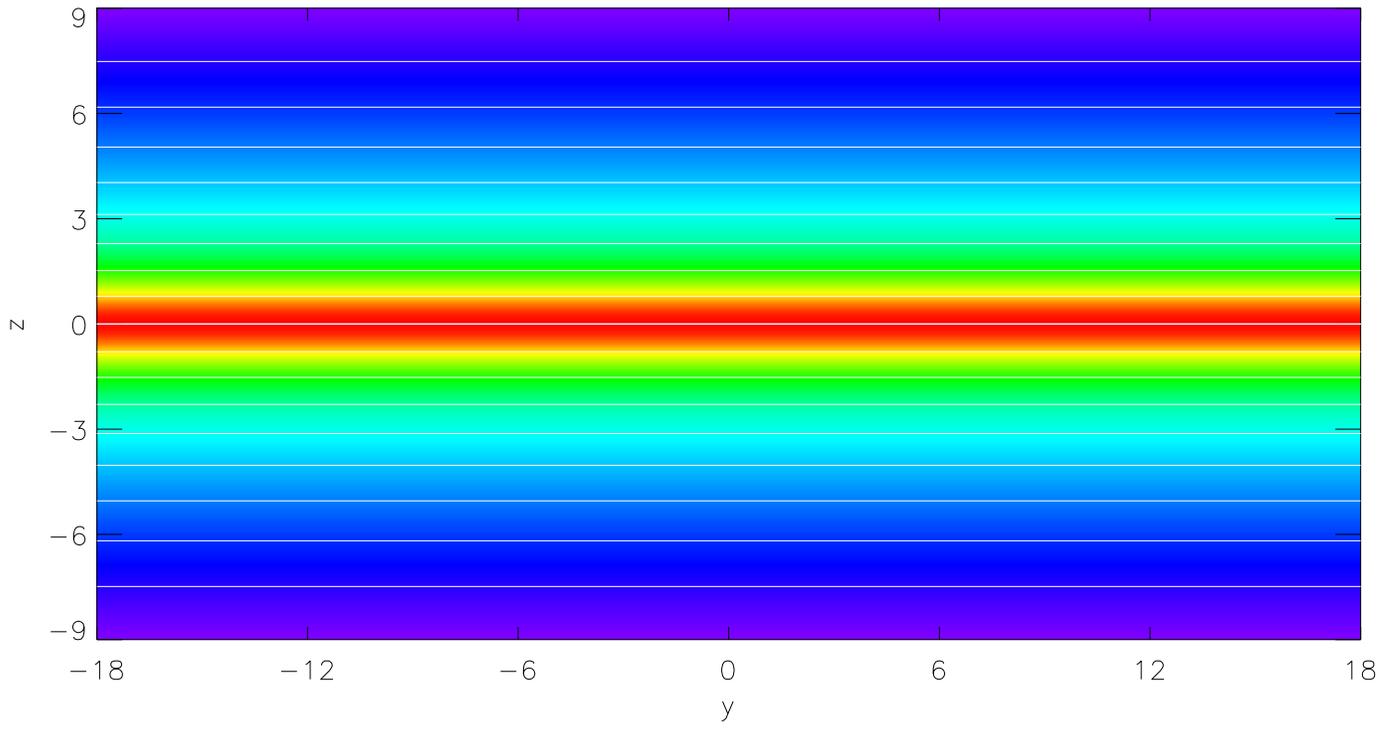
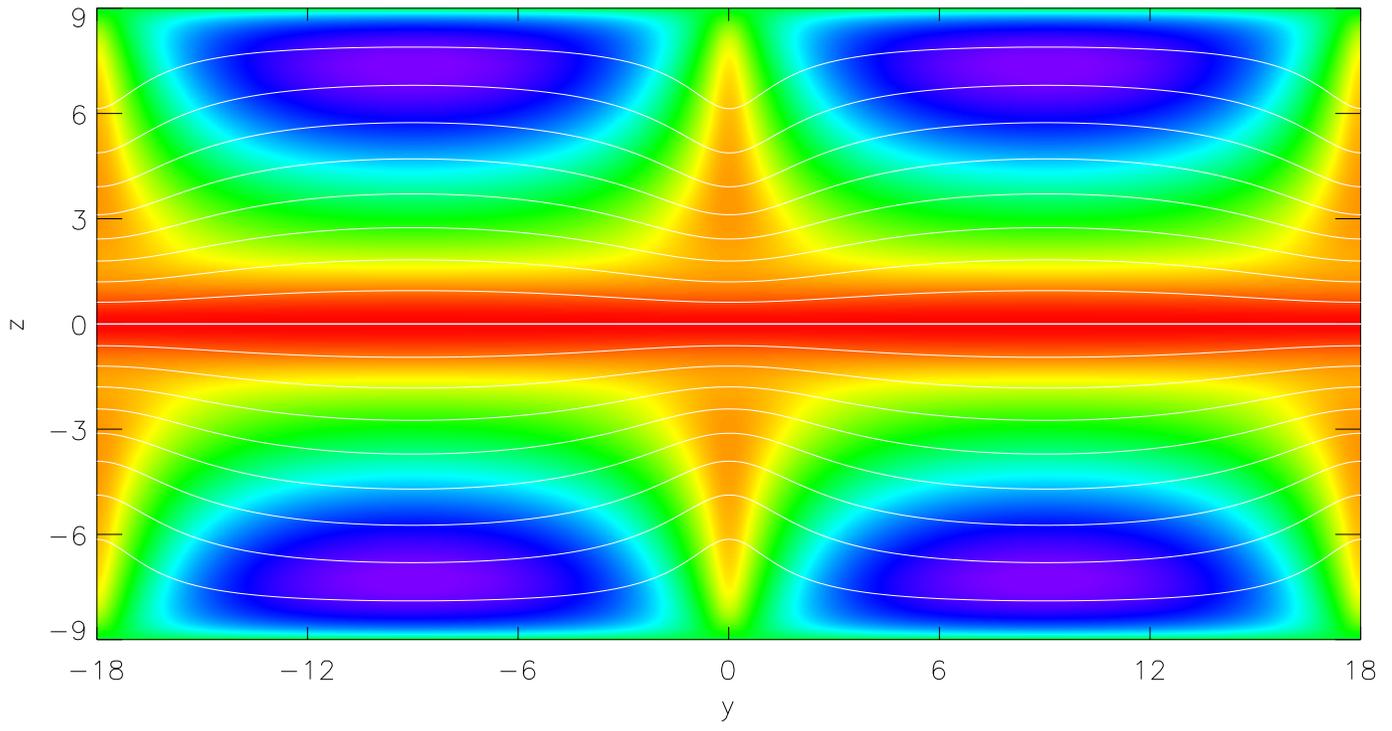
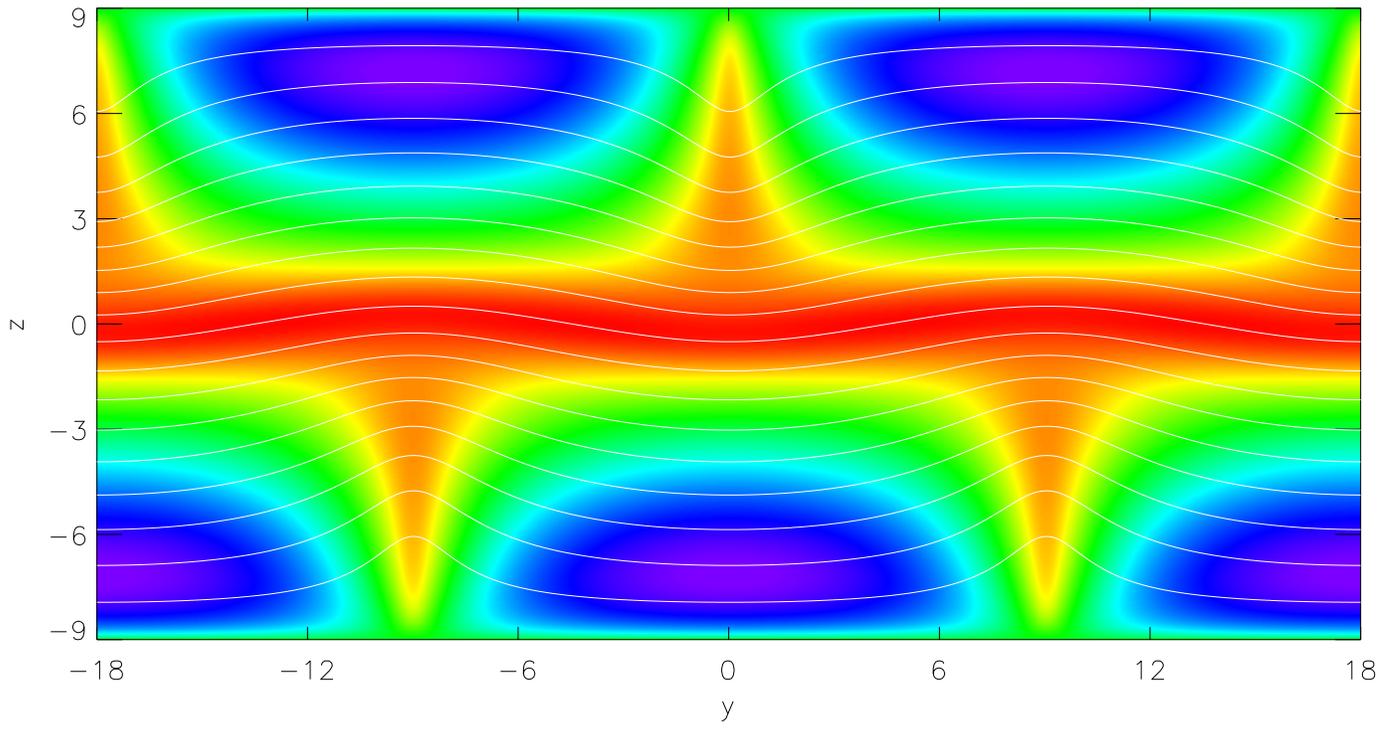
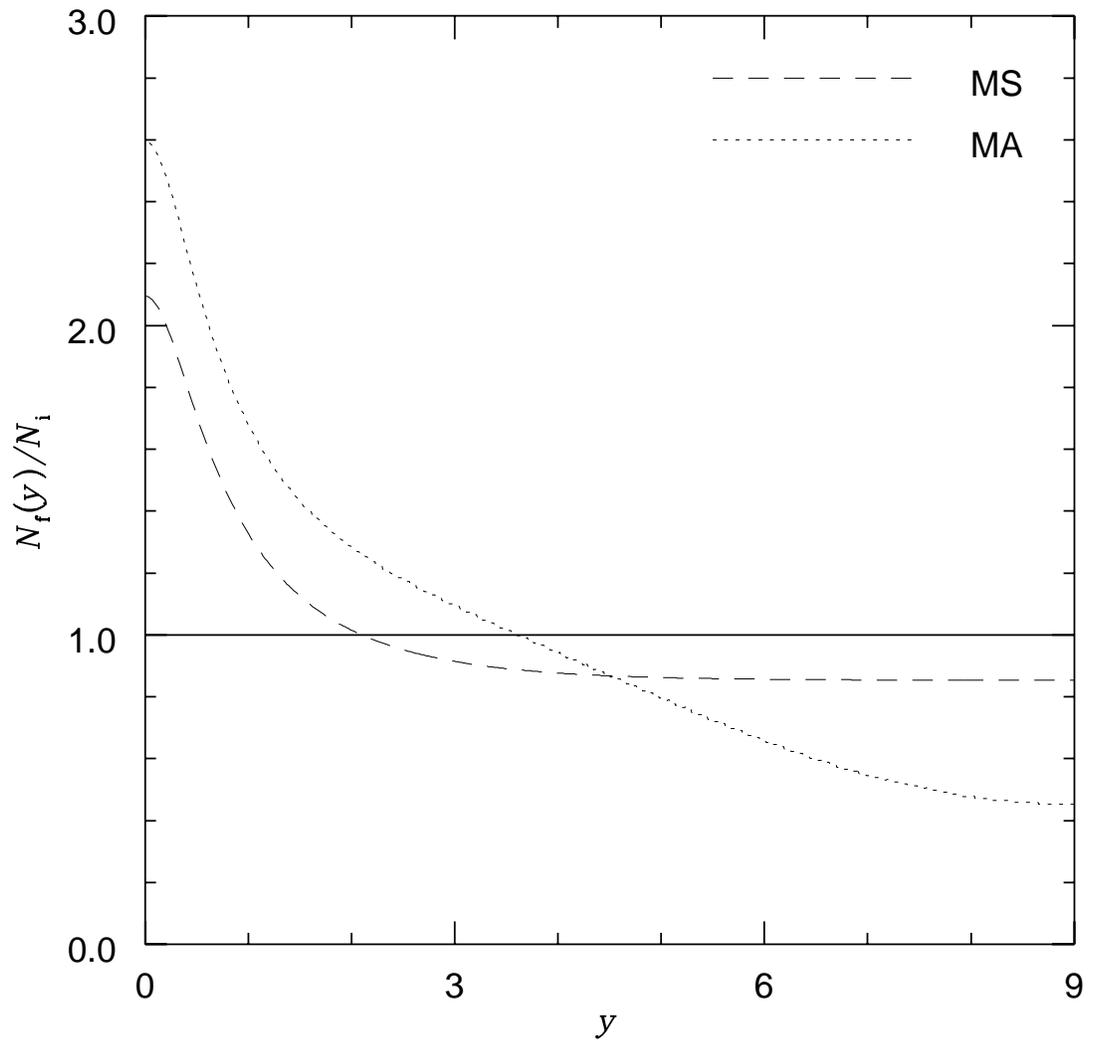
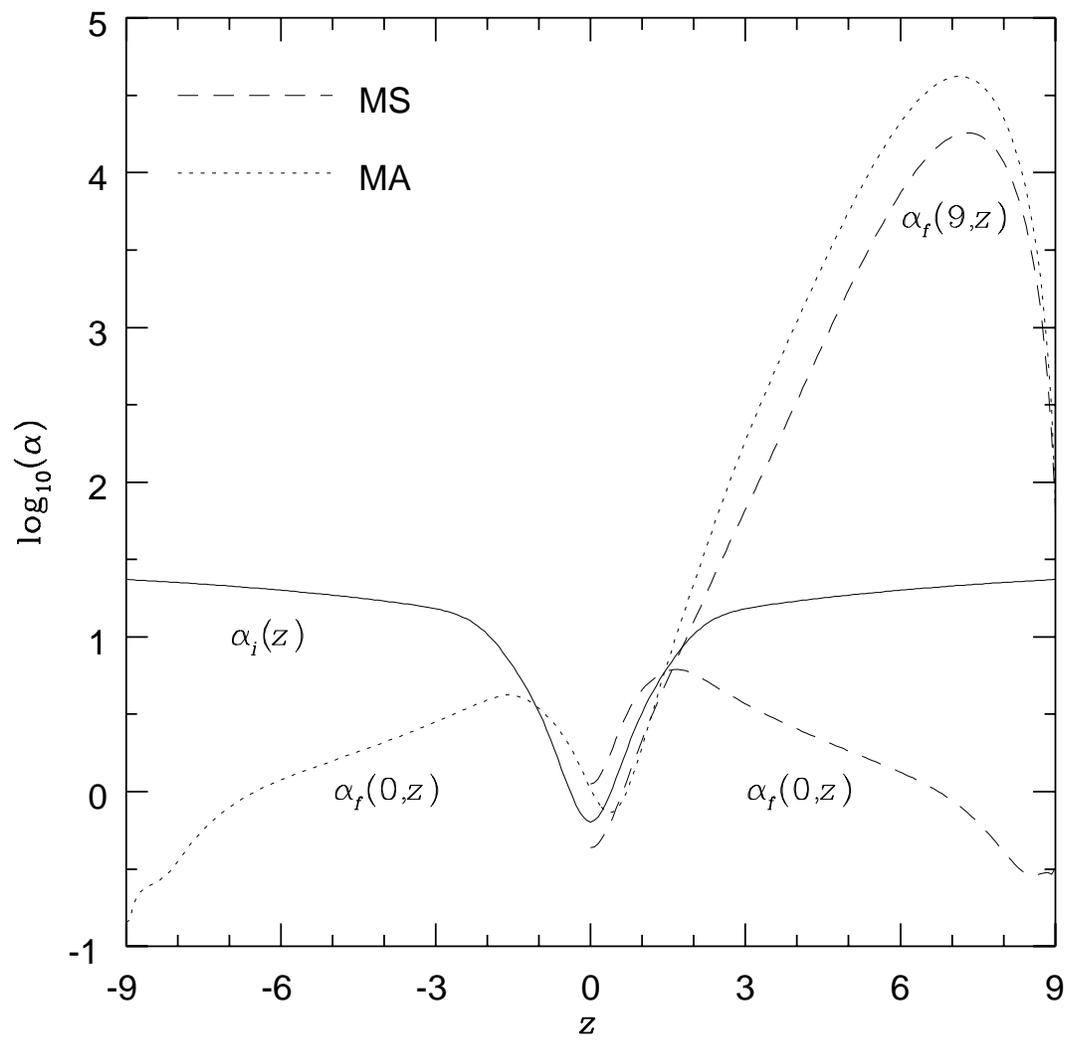
\end{document}